\baselineskip=18pt plus 2pt minus 1pt
\magnification=1100
\hsize=6.0truein
\vsize=8.4truein
\voffset=24pt
\hoffset=.1in

\centerline{\bf Explanation of Giant Cluster Coexistence }
\centerline{\bf In Doped Manganites and Other Compounds}

 \vskip 0.8cm
\centerline{Adriana Moreo, Matthias Mayr, Adrian Feiguin,
Seiji Yunoki, and Elbio Dagotto}
\vskip.15in
\centerline{\it National High Magnetic Field Lab and Department of Physics,}
\centerline{\it Florida State University, Tallahassee, FL 32306, USA}
\centerline{(Submitted 12 November 1999)}
\vskip .15in

\centerline{{\bf Abstract}}

Computational studies show the generation of 
large coexisting metallic and insulating clusters with
 equal electronic density in models for manganites.
The clusters are induced by disorder on exchange and
hopping amplitudes near
first-order transitions of the non-disordered strongly coupled system.
The random-field Ising model is used to explain the qualitative aspects of
our results. Percolative characteristics
are natural in this context.
Our results explain
the recently experimentally
discovered micrometer size inhomogeneities in manganites.
The conclusions are general and apply to a variety of compounds.

\vskip 1.0cm


Considerable work is currently being focussed on the experimental and
theoretical study of manganese oxides.
This huge effort was triggered by 
the discovery of Colossal
Magnetoresistance (CMR) in manganites (1),
where the resistivity changes
 by several orders of magnitude upon the
application of modest fields of a few Tesla
at the carrier densities and temperatures where metallic and insulating
phases are in competition. 
Theoretical investigations of simple manganite models
 based upon numerical simulations and mean-field approximations
have reproduced some
of the complex spin-, orbital- and charge-ordered phases observed
experimentally. In particular, the
charge-ordered (CO) CE-state of half-doped manganites 
has been recently stabilized in Monte Carlo (MC) simulations of the 
two-orbital model with Jahn-Teller phonons (2). This state is in 
competition with the ferromagnetic (FM) and A-type states 
also observed in experiments at the hole density x=0.5. However, the curious
magnetotransport properties of manganites has resisted theoretical
understanding and a proper
explanation of the CMR phenomenon is still lacking.

Recently, important new experimental information about the
microscopic properties of manganites has been reported.
Using electronic diffraction and transport
techniques, Uehara {\it et al.} have investigated the effect of
Pr-doping upon the metallic ferromagnetic compound $\rm La_{5/8} Ca_{3/8} Mn
O_3$ (3). As the Pr density (y) increases by the replacement
$\rm La$$\rightarrow$$\rm Pr$, the system 
changes at y$\sim$0.35 to an insulating CO-state.
At low temperature in this regime the unexpected 
coexistence of $giant$ clusters
of FM and CO phases was observed (3). 
Similar results were reported by F\"ath {\it et al.} using
scanning tunneling spectroscopy applied to $\rm La_{0.7} Ca_{0.3} Mn
O_3$ (4). At temperatures 
close to the Curie temperature, thin-film spectroscopic
images revealed a complicated pattern of interpenetrating giant metallic and
insulating phases.
The clusters found in both these experiments
were as large as 0.1 $\mu$m=1000$\rm \AA$$\sim$250$a$, with
$a$$\sim$4$\rm \AA$ the Mn-Mn distance (3,4). 
The metal-insulator FM-CO transition occurs
through a percolative process among the clusters, as a function of
either temperature or magnetic field.
These results rule out the picture of homogeneously
distributed small polarons to describe doped manganites in the CMR regime.

The discovery of
 huge coexisting FM-CO clusters in a manganite single-crystal
is puzzling.
The only theoretical framework which in principle could
be used to address this issue is the phase separation (PS) scenario
where mixed-phase characteristics, involving phases with different
electronic densities, are natural (5). 
The PS ideas are indeed successful in describing
manganites at, e.g., small hole density, where nanometer size
inhomogeneities have been widely discussed (5), and at high
densities x$\sim$1 based on recent magnetic and transport data (6).
However, 
the micrometer clusters at intermediate densities found in Refs.(3,4)
appear to require an alternative explanation since the energy cost of
charged $\mu$m-size domains would be too large to keep the structure
stable.
Actually, explicit numerical calculations in one dimensional (1D)
models have shown that the large clusters in PS regimes
break down into smaller pieces of a few lattice spacings in size
 upon the introduction of a
nearest-neighbor charge repulsion (7).
In addition,
this repulsion tends to arrange the charge in an ordered
pattern (7) --charge-density-waves or stripes-- 
contrary to the random location and
shape of the clusters observed experimentally (3,4). 
A novel framework involving large clusters with $equal$-$density$ phases is needed to
rationalize the results of Refs.(3,4).

In the absence of theoretical proposals to explain the
giant FM-CO clusters it is
necessary  to reconsider some of the properties of the models 
studied thus far. Of particular relevance is the assumption of 
translationally invariant interactions, which is at odds with the
chaotic looking appearance of microclusters in experiments (3,4). For this
reason, here we report a
computational study of manganite models which combines (i)
strong coupling interactions, necessary to reproduce 
the rich variety of 
ordered phases of these materials, and (ii) quenched disorder.
The latter is caused by
the random chemical replacement of ions, such as La and Pr, with
different ionic sizes. This replacement affects the hopping
amplitudes of $\rm e_g$-electrons due to the buckling of the Mn-O-Mn
bonds near Pr (8). Recent calculations showed 
that the concomitant
modification of the exchange coupling $\rm J_{AF}$
among  the $\rm t_{2g}$-spins is likely equally important in establishing
the properties of manganite models (2). Thus, by considering randomly chosen 
hopping and exchange couplings 
fluctuating about the non-disordered values
of interest, the physics of doped manganites will be more properly
captured. Following this
procedure, here we report the natural appearance
of coexisting giant clusters of equal-density FM and AF
phases in realistic models. The
conclusions are general and similar cluster formation is expected
for a variety of compounds.


To present our main results first consider
the 
two-orbital model, described extensively in
previous work (5,9). 
It contains (i) an electronic hopping term, regulated by amplitudes $\rm
t^{\alpha}_{ab}$, with a,b=1,2 labeling the 
$\rm d_{x^2 -y^2}$ and $\rm d_{3z^2 - r^2}$ orbitals, $\alpha$=x,y,z
being the axes directions,
and $\rm t^{x}_{11}$=t the energy scale, (ii) a strong
FM 
coupling between the localized $\rm t_{2g}$- and mobile $\rm e_g$-fermions,
regulated by $\rm J_H$, 
(iii) a direct antiferromagnetic (AF) exchange among the
localized spins with strength $\rm J_{AF}$, and (iv) an electron-phonon
coupling between the $\rm Q_2$ and $\rm Q_3$ Jahn-Teller modes and the
mobile electrons, with strength $\lambda$. The phase diagram of the
non-disordered
model was studied by standard MC simulations
using classical localized spins and phonons (2,9). Similar results  
were obtained with mean-field approximations 
including Coulombic repulsions (2). The generality and rationalization
of the numerical data described below suggest that the main conclusions 
are actually independent of the detailed properties of the competing
 states. Whether the phases 
are generated by phononic, magnetic, or Coulombic interactions appears unimportant.

The focus of our studies will be on $first$-$order$ transitions, which
in the two-orbital model
occur in several locations in parameter space in any dimension of interest
(2,5,9). However, for the
disorder-induced  cluster formation described below 
it is more convenient
to analyze 1D systems first since the two-orbital 2D lattices that can be studied
computationally are not sufficiently large. 
Among the possible 1D first-order
transitions, results are here reported for the transition occurring between
FM and AF states at
fixed x=0.5
and large $\lambda$, as a function of $\rm J_{AF}/t$. 
The AF phase studied has a four-spin 
unit cell $\uparrow$$\uparrow$$\downarrow$$\downarrow$,
and a concomitant peak in the spin structure factor
S(q) at q=$\pi/2$ (2,10). The state is insulating, as demonstrated by
the absence of Drude weight and the vanishing density of states
at the Fermi energy. 
Nearest-neighbors correlations among the $\rm t_{2g}$-spins
are used to distinguish among the FM and AF phases. 
In Fig.1A the energy per site 
(E) vs $\rm J_{AF}/t$ for the non-disordered model is shown. 
The dE/d$\rm (J_{AF}/t)$ discontinuity indicates the first-order
character of the transition at $\rm J_{AF}/t|_c$$\approx$0.21. 
Disorder is introduced in 
$\rm t^{\alpha}_{ab}$ and  $\rm J_{AF}$ such that 
$\rm J_{AF}/t$ becomes effectively random
in the interval $\rm J_{AF}/t|_c$-$\delta$ to
$\rm J_{AF}/t|_c$+$\delta$. Results for one fixed set of couplings are shown 
for $\delta$=0.01 in Fig.1B (other sets lead to similar results).
The MC averaged correlations in Fig.1B already 
show one of the main results of this paper,
namely the remarkable formation of coexisting large FM and AF
 clusters in the ground state, 
typically of order 10$a$ each ($a$ is the lattice spacing). 
This occurs even though
$\rm J_{AF}/t$ at each link (not shown)
rapidly changes at the $a$ scale since different
sites are uncorrelated in the disorder. Naively it may have been expected that 
at every link either the FM or AF phases would be stable depending on
the value of $\rm J_{AF}/t$, as it occurs for a dominant strong disorder.
However, at weak disorder
this would produce a large 
interface energy and the order parameter cannot
follow the rapid oscillations of $\rm J_{AF}/t$ from site to site. As a 
consequence, structures much larger than the lattice spacing 
emerge, with a size 
regulated by $\delta$ (for instance, in Fig.1C results at $\delta$=0.05
contain FM clusters smaller than in Fig.1B).
The effect occurs only near first-order
transitions, i.e. the same weak disorder in other regions does not produce
important effects in the spin correlations.
Qualitatively
similar results appear also in other first-order
transitions of the two-orbital model, such as for the
FM-CO(CE-state) level crossing reported in Ref.(2) using
4$\times$4 and 4$\times$4$\times$2 clusters. 
The generation of large equal-density
clusters by ($\rm t,J_{AF}$)-disorder near first-order transitions
is an effect unforeseen in previous manganite
investigations. 


The rapid CPU time growth with
cluster size of the two-orbital model  does not allow us to investigate
numerically the phenomenon in more detail than shown in Figs.1A-C. 
Fortunately, there are simpler models with the same behavior, including 
the well-known one-orbital model (5). It
contains hopping for only one species of $\rm e_g$-electrons (regulated
by t), a FM Hund coupling $\rm J_H$ linking the
$\rm e_g$- and (classical) $\rm t_{2g}$-spins, 
and  a direct exchange $\rm J_{AF}$
among the $\rm t_{2g}$-spins.
Previous work showed that this
model also has a first-order transition at 
x=0.5 as $\rm J_{AF}/t$ varies, in the large $\rm J_H$ regime (10). 
It involves equal-density metallic FM and
insulating AF states, the latter with a similar spin structure as
the AF state of Figs.1A-C.
To investigate disorder effects here the natural modification 
is to select
the exchange $\rm J_{AF}$ randomly in the interval [$\rm
J_{AF}^c$-$\delta$,$\rm J_{AF}^c$+$\delta$] (11), where 
$\rm J_{AF}^c$$\sim$0.14 is the critical
first-order transition coupling at $\rm
J_H$=$\infty$, t=1, and T=1/70 in the non-disordered limit
(Fig.1D). 
In Fig.1E, results of a
MC simulation corresponding to a representative set of random couplings 
$\rm
(t,J_{AF})$ centered at 0.14
are shown. As in the two-orbital case, 
FM and AF clusters, this time as large as 20$a$,
are easily obtained.
S(q) (not shown) contains a double-peak structure with
dominant features indicating a FM-AF mixed phase.
If the range of possible  $\rm (t,J_{AF})$ increases,
the cluster size decreases (Fig.1F).
Open boundary conditions (OBC) were used in Figs.1D-F, and
periodic boundary conditions (PBC) in Figs.1A-C, to show that the
large cluster formation occurs independently of these details.


To further investigate the universality of the large cluster generation
phenomenon,
note that the non-disordered one-orbital model
has another prominent first-order transition
corresponding to a discontinuity
in the density $\rm \langle n \rangle$ 
vs chemical potential $\mu$, for a wide
range of couplings (5). The direct interpretation of such a result
is the presence of PS between competing FM and AF states (5).
However, in the context emphasized here
the focus shifts from the transition properties to
the effect of disorder on 
the $\rm \langle n \rangle$ discontinuity itself. Disorder is
 here naturally introduced as a
site-dependent chemical potential of 
the form $\rm \sum_{\bf i} \phi_{\bf i} n_{\bf i}$, where
$\rm \phi_{\bf i}$ is randomly selected in the interval 
[-$\rm {{W}\over{2}}$,+$\rm {{W}\over{2}}$], and 
$\rm n_{\bf i}$ is the electronic number operator at site $\rm {\bf i}$.
Results of a standard MC simulation 
for a L=20 sites chain of the disordered one-orbital model
 are in Fig.2A. Averages of $\rm \langle n \rangle$ over
100 disorder configurations are shown.
For the values of W
studied here, $\rm \langle$n$\rm \rangle$ no longer has a discontinuity.
For small W the
first-order transition is replaced by a rapid crossover, where the
compressibility proportional to $\rm d\mu/d\langle n \rangle$ 
remains high, suggesting the formation of large clusters.
This is confirmed in Fig.2B where the 
nearest-neighbor $\rm t_{2g}$-spin correlations are
shown for a L=60 chain, two disorder configurations,
 and one (typical) MC snapshot for each. These results are
very similar if averages over the MC configurations are made, showing
that the system is basically 
frozen into an inhomogeneous ground state with
large AF and FM clusters of size $\sim$10-20$a$,
similarly as in Fig.1B,E. 
Once again, reducing W increases the cluster
sizes, only limited by the size of the systems that
can be studied computationally.
Here it was also observed that the mixed-phase ground state
leads to a $pseudogap$ in the 
T$\sim$0 density of states (Fig.2C),
as it occurs in non-disordered models
at finite temperatures in particular regions of parameter space (12).
An analogous pseudogap was also observed working with the
two-orbital model. 
Similar results as in Figs.2A-C 
were also found in two dimensions (2D) 
(see, e.g., Fig.2D) and thus the large cluster formation
 certainly does not depend on  
pathological properties of 1D systems.
Note that in the particular example studied in
Figs.2A-D, the AF-FM regions involved have different electronic densities,
complementing the results of Figs.1A-F with equal-density clusters. In
both cases the results are illustrative of cluster formation induced 
by disorder.


The simplicity and universality of 
the MC simulation results with large coexisting
FM-AF clusters suggest that there is
a general principle at work in the problem. To understand this effect
let us briefly review the phenomenology 
of the random-field Ising model (RFIM) (13)
defined by the Hamiltonian
$\rm H = - J \sum_{\langle {\bf ij} \rangle} S_{\bf i} S_{\bf j} - 
\sum_{\bf i} h_{\bf i} S_{\bf i}$,
where $\rm S_{\bf i}$=$\rm \pm 1$, and the rest of the notation is standard.
The random fields $\rm \{ h_{\bf i} \}$ 
have the properties $\rm [h_{\bf i} ]_{av}$=0 and 
$\rm [h^2_{\bf i} ]_{av}$=$\rm h^2$, where h characterizes the width of the
distribution, and 
$\rm [...]_{av}$ is the average over the fields.
In manganites 
the Ising variables represent the competing metallic and
insulating states on a small region of space centered at $\rm {\bf i}$. 
The random field mimics the $\rm t^{\alpha}_{ab}$
and $\rm J_{AF}$ fluctuations
locally favoring one state over the other. 
Without disorder, the Ising model has a first-order 
transition at zero magnetic field and T=0 between the
two fully-ordered states,
analogous to the AF-FM first-order transitions 
of non-disordered manganite models (2). However, at $\rm h \neq 0$ 
the properties of the Ising
transition are drastically affected (13).
The key arguments guiding RFIM investigations (14)
can be restated for
manganites. Working very close to a first-order transition,
consider that in a region dominated by phase-I (either AF or FM),
a phase-II bubble of radius R is created. The
energy cost $\rm R^{d-1}$
 is proportional to the domain wall area, with
d the spatial dimension. To stabilize
the bubble it is necessary to induce an energy compensation 
originated in the 
$\rm (t, J_{AF})$ disorder.  Consider
the average hopping
 inside the bubble using $\rm S_R$=$\rm \sum_l t_l$, where l labels bonds and $\rm
t_l$ is the hopping deviation at bond l
from its non-disordered value,
the latter of which is fixed at the critical coupling of
the first-order transition of the non-disordered model.
Although  
the random hopping deviations mostly cancel inside
the bubble, important fluctuations must be considered. 
In particular, the standard deviation of $\rm S_R$  is
$\rm \sigma_{S_R}$=$\rm (\Delta t) R^{d/2}$ since  $\rm
[t_{l} t_{l'}]_{av}$=$\rm (\Delta t)^2 \delta_{ll'}$, 
with $\rm (\Delta t)$ characterizing the width of the random
hopping distribution about the non-disordered value. 
A similar expression holds for the $\rm J_{AF}$ fluctuations. 
Then, centered at any lattice site it is always possible
 to find a region of size R, such that at least
the average couplings
favor either phase-I or -II with a substantial 
strength of order $\rm
R^{d/2}$, although 
individual random deviations $\rm
t_l$ cannot exceed a (small number) $\rm \Delta t$.

To illustrate the generation of large clusters
in the RFIM, standard MC simulations were performed. Although similar 
MC studies
have been discussed in the RFIM framework (13), 
the results shown here provide useful qualitative
information to manganite experts. 
In Fig.3A, low temperature results are shown for
one representative set $\rm \{ h_{\bf i} \}$ individually taken from
$\rm [-W,+W]$
with W=3.0, in units of
J=1 (W=$\sqrt{3}$h). The dynamical formation of large coexisting
 clusters 
is clear, in spite of the uncorrelated character of the  random fields in
neighboring sites. Using the same set 
$\rm \{ h_{\bf i} \}$ as in Fig.3A but
rescaling its intensity with W,
Fig.3B shows that as W is reduced the
typical cluster sizes rapidly grow and at W=1.5 clusters as large
as 50$a$ in characteristic length are possible. 
Fig.3C contains simulation results now
on a large
500$\times$500 lattice showing that RFIM cluster sizes
can be made as large as those
found in manganite experiments (250$a$) by simply adjusting W.
Fig.3D illustrates the influence of an external
field -$\rm H_{ext} \sum_i S_{\bf i}$ added to the Hamiltonian. As
$\rm H_{ext}$ grows the region most affected by the field is the
surface of the spin down domains, which are transformed
into spin up. This tends to suppress the narrowest regions of the spin down
clusters, as highlighted with arrows in Fig.3D, 
providing a field-induced connection among spin up
regions that otherwise would be disconnected. Then,
intuitively, as $\rm H_{ext}$ increases a percolative
transition is to be expected. 
Based on the RFIM-manganite
analogy, the picture described here predicts a similar
percolative transition involving metallic and insulating clusters
as chemical compositions, temperatures, or magnetic
fields are varied near first-order transitions, as
 observed experimentally (3,4).
Giant cluster generation by weak disorder in manganite
models and in the RFIM appear related phenomena.
However, at this early stage in the calculations it is
difficult to predict critical exponents for the metal-insulator
transition. Even for simpler
spin systems such as diluted anisotropic antiferromagnets
there is still no full agreement between RFIM theory and experiments (13). 
In addition,
the manganite critical dimension may be affected by the
1D character of the zig-zag chains that form the planar
CE-state (15), and 
critical slowing down as in the RFIM can produce
rounding effects that make a comparison between scaling theory and
manganite experiments difficult.


The ideas described here are not limited to particular manganite
compounds but they apply to other materials where a 
transition with first-order characteristics occurs, either 
by varying temperatures in compounds with some source of disorder,
or by explicit chemical substitution which
leads to quenched fluctuations in the hopping and exchange
amplitudes. For instance, other manganites such as 
$\rm (La_{1-x} Tb_x )_{2/3} Ca_{1/3} Mn O_3$ 
and 
$\rm La_{2-2x}Sr_{1+2x} Mn_2 O_7$ also have a AF-FM competition 
at low temperatures. While previous investigations assigned 
spin-glass (16) or canted-phase (17) 
characteristics to the intermediate region,
mixed-phase properties involving equal-density large clusters 
as found in 
$\rm La_{5/8-y} Pr_y Ca_{3/8} Mn O_3$ 
provide an
alternative description. In $\rm La_{0.5} Ca_{0.5} Mn O_3$,
large FM and CO clusters have also been reported (18), and the influence of disorder
on the first-order transition is a possible explanation for
their existence (19). The concepts
discussed here also apply   to ruthenates, such as 
$\rm (Sr_{1-y} Ca_y)_3 Ru_2 O_7$, where a 
difficult-to-characterize
$\rm y$$\sim$0.5 region separates FM and AF phases (20). 
The metal-insulator transition of $\rm La Ni_{1-x} Fe_x O_3$ (21) may also
proceed through a mixed-phase (equal-density) regime with giant cluster formation.
In addition, $\rm Eu B_{6}$ behaves similarly to manganites 
(22) and it may present an analogous percolative behavior caused
by disorder. The same could occur for 
the transition metal chalcogenide $\rm Ni S_{2-x} Se_x$ at x$\sim$0.5 (23).
Cr alloys such as $\rm Cr_{1-x} Fe_x$, 
may also have an interesting AF-FM
competition with percolative properties.
Finally, the notorious inhomogeneities observed experimentally
in high temperature superconductors,
such as $\rm La_{2-x} Sr_x Cu O_4$, in regimes
of low density of carriers may be caused in part by
disordering effects on first-order transitions.


Summarizing, based on the calculations reported here,
the giant clusters in manganites found experimentally in Refs.(3,4)
are conjectured to
be caused by quenched disorder in the couplings ($\rm t^{\alpha}_{ab}$
and $\rm J_{AF}$) of the system, which are induced by chemical
substitution, and which affect transitions that otherwise would be of
first-order without disorder (24). 
A pseudogap in the density of states
appears in this regime. The mixed-phase state 
presented in Figs.1A-F involves clusters with
equal electronic density, complementing the phase separation
scenario which involves regions with different
densities (5). In phase-separated regimes 
disorder also leads to cluster formation (Figs.2A-D).
Although non-disordered models remain crucial to determining the
competing tendencies in manganites and to establish the order of
the phase transitions, disordering effects appear necessary to
reproduce the subtle percolative nature of the metal-insulator transition and
the conspicuous presence of $\mu$m domains in these compounds in regimes
near first-order transitions (3,4).
The present observations are
general, not based on fine tuning of models or parameters, and 
they should apply to a variety of other compounds as well.
The formation of coexisting giant clusters when 
two states are in competition through first-order transitions
should be a phenomenon frequently present in 
transition-metal-oxides and related compounds.



\vskip 2.0cm

\centerline{REFERENCES}
\medskip

\item{1.} 
Y. Tokura {\it et al.}, J. Appl. Phys. {\bf 79}, 5288 (1996);
A. P. Ramirez,  J. Phys.: Condens. Matter {\bf 9}, 8171 (1997).

\item{2.} S. Yunoki, T. Hotta,
and E. Dagotto, preprint, cond-mat/9909254.

\item{3.} M. Uehara, S. Mori, C. H. Chen, and S.-W. Cheong, 
Nature {\bf 399}, 560 (1999).

\item{4.} M. F\"ath, S. Freisem, A. A. Menovsky, Y. Tomioka, J. Aarts,
and J. A. Mydosh, Science {\bf 285}, 1540 (1999).

\item{5.} A. Moreo, S. Yunoki and E. Dagotto, Science {\bf 283},
2034 (1999); and references therein.

\item{6.} J. J. Neumeier and J. L. Cohn, preprint.

\item{7.} A. L. Malvezzi, 
S. Yunoki, and E. Dagotto,  Phys. Rev. B{\bf 59},
7033 (1999).

\item{8.} S.-W. Cheong, and H. Y. Hwang, in {\it Colossal Magnetoresistance
Oxides}, ed. Y. Tokura, Monogr. in Condensed Matter Sci., 
Gordon \& Breach, London, 1999.

\item{9.} S. Yunoki, A. Moreo,
and E. Dagotto, Phys. Rev. Lett. {\bf 81}, 5612 (1998). See also
A. J. Millis, B. Shraiman and R. Mueller, Phys. Rev.
Lett. {\bf 77}, 175 (1996). 

\item{10.} S. Yunoki and A. Moreo, Phys. Rev. B{\bf 58}, 6403 (1998).

\item{11.} Simulations with other distributions of
random numbers lead to similar results.

\item{12.} A. Moreo, S. Yunoki, and E. Dagotto, Phys. Rev. Lett. {\bf
83}, 2773 (1999).

\item{13.} See contributions by T. Natterman (cond-mat/9705295)
and D. P. Belanger (cond-mat/9706042) to {\it Spin Glasses and Random  Fields}, ed.
A. P. Young, World Scientific.

\item{14.} Y. Imry and S. K. Ma, Phys. Rev. Lett. {\bf 35}, 1399 (1975).
See also M. Aizenman and J. Wehr, Phys. Rev. Lett. {\bf 62},
2503 (1989).

\item{15.} T. Hotta {\it et al.}, preprint.

\item{16.} J. M. De Teresa {\it et al.}, Phys. Rev. B{\bf 56},
3317 (1997).

\item{17.} M. Kubota {\it et al.}, cond-mat/9902288.

\item{18.} S. Mori, C. H. Chen, and S.-W. Cheong, Phys. Rev. Lett. 
{\bf 81}, 3972 (1998).

\item{19.} The results of this paper are also related to the
relaxor FM picture of Cr-doped $\rm Nd_{1/2} Ca_{1/2} Mn O_3$
very recently 
discussed in T. Kimura, Y. Tomioka, R. Kumai, Y. Okimoto, and Y. Tokura,
Phys. Rev. Lett. {\bf 83}, 3940 (1999).

\item{20.} G. Cao {\it et al.}, Phys. Rev. B{\bf 56}, 5387 (1997).

\item{21.} D. D. Sarma {\it et al.}, Phys. Rev. Lett. {\bf 80},
4004 (1998). These authors  
observed that the usual paradigms 
are not enough to describe
$\rm La Ni_{1-x} M_x O_3$ (M=Mn,Fe),
and they anticipated that disorder plays an important
role in this compound.

\item{22.} S. Yoon {\it et al.}, Phys. Rev. B{\bf 58}, 2795 (1998);
J. L. Gavilano {\it et al.}, Phys. Rev. Lett. {\bf 81}, 5648 (1998);
S. L. Cooper, private communication.

\item{23.} A. Husmann {\it et al.}, Science {\bf 274}, 1874 (1996).

\item{24.} The effects described in this paper are
different from other proposals for manganites
where disorder leads to 
carrier localization.
In such a  context the formation of giant clusters is unnatural,
and first-order transitions 
in non-disordered models are not needed for the localization
to occur. The large static clusters found here must also be distinguished 
from the dynamical small cluster
formation at finite temperature in non-disordered models, with
clusters evolving rapidly with MC time, changing shapes and sizes 
such that the translational invariance is restored when time-averaged (5). On
the other hand, the disorder described here pins the cluster at
particular locations, it involves phases with the same density,
 and the typical sizes are much larger than
found in previous simulations (5).
Finally, note that the clusters in Figs.1-3 
should also not be confused with
metastable states arising from the first-order character of the
transition in non-disordered models.

\item{25.} The authors thank W. Bao,
 S.-W. Cheong,  V. Dobrosavljevi\'c, T. Hotta,
and K. Yang for very useful discussions. A.M. and E.D. 
are supported in part by grant NSF-DMR-9814350.
A.F. thanks the Fundacion Antorchas for partial support.

\vskip 0.7cm
\centerline{\bf Figure Captions}
\vskip 0.4cm

\item{1.} (A-C) are MC results for the two-orbital model with
$\rm \langle n \rangle$=0.5, T=1/100, $\rm J_H$=$\infty$, $\lambda$=1.2, t=1,
PBC,
and L=20 (chain size). (A) is the energy per site vs $\rm J_{AF}/t$ for
the non-disordered model.
A level crossing (first-order transition) 
between FM and AF states occurs; (B) 
MC averaged nearest-neighbor
$\rm t_{2g}$-spins correlations 
vs position along the chain for one set of
random $\rm t^{\alpha}_{ab}$ and $\rm J_{AF}$ couplings
such that $\rm J_{AF}/t$ at every site lies
between 0.21-$\delta$ and 0.21+$\delta$,
with $\delta$=0.01. FM and AF regions are highlighted; 
(C) Same as (B) but using $\delta$=0.05;
(D-F) are results for the one-orbital model with 
$\rm \langle n \rangle$=0.5, T=1/70, $\rm J_H$=$\infty$, t=1, OBC, 
and L=64. (D) is the energy per site vs 
$\rm J_{AF}$ for the non-disordered model, 
showing the level crossing 
between FM and AF states at $\rm
J_{AF}$$\sim$0.14; (E) are the MC averaged nearest-neighbor 
$\rm t_{2g}$-spin correlations vs position for one distribution of
random hoppings and $\rm t_{2g}$ exchanges such that $\rm J_{AF}/t$ is
now distributed  between 0.14-$\delta$ and 0.14+$\delta$,
with $\delta$=0.01; 
(F) Same as (E) but with $\delta$=0.03.

\item{2.} Results of a MC simulation of the one-orbital model
with a random chemical potential, PBC, 
$\rm J_H$=8.0, and $\rm J_{AF}$=0.0, in units of t=1.
(A) $\rm \langle n \rangle$ vs $\mu$ for a L=20 chain at
 T=1/75 using 24,000 MC sweeps per $\{ \phi_i \}$ set. 
The results are averages over $\sim$100 of $\{ \phi_i \}$
configurations for the values of W shown; (B)
Nearest-neighbors $\rm t_{2g}$-spin correlations vs
their location along a L=60 chain with $\rm \mu$=-6.7 and T=1/75. Shown are results
for one representative
MC snapshot, W=0.25 (upper panel) and W=1.0 (lower panel).
Other snapshots differ from this one only 
by small fluctuations. The FM-AF clusters remain pinned at the same
locations as the simulation evolves; (C) Density of states 
at T=1/75, L=20 and $\rm
\mu$=-6.7 showing the presence of a pseudogap.
The average density is $\rm \langle n \rangle$$\sim$0.87; 
(D) Results of a representative
MC snapshot for an 8$\times$8 cluster, T=1/50,
$\mu$=-6.2 (close to the critical value), 
and W=1.0. Regions with FM or AF 
nearest-neighbor $\rm t_{2g}$-spin correlations are shown.

\item{3.} Results of a MC simulation of the 2D RFIM at T=0.4 (J=1), with PBC. 
The dark (white) 
small squares represent spins up (down). At T=0.4 the
thermal fluctuations appear negligible and the results shown are those
of the lowest energy configuration.
(A) was obtained
for  W=3, $\rm H_{ext}$=0  using a 100$\times$100 cluster and
one set of random fields $\rm \{ h_{\bf i} \}$.
Typical cluster sizes are $\sim$10$a$; (B) Same as (A) 
but with W=1.5. The cluster sizes have grown to $\sim$50$a$; 
(C) Results using
a 500$\times$500 cluster with W=1.2, $\rm H_{ext}$=0 and for one
configuration of random fields. The giant and percolative-like features of the
clusters are apparent in the figure;
(D) Same as (C) 
but now contrasting results between zero and nonzero $\rm H_{ext}$.
The dark regions are spins up in the $\rm H_{ext}$=0 case, the grey
regions are spins down at zero field that have flipped to up at $\rm
H_{ext}$=0.16, while the white regions have spins down with and without
the field. Special places are arrow marked
where narrow spin down regions have flipped linking
spin up domains.
In (A)-(B) and (C)-(D) the same set $\rm \{ h_{\bf i} \}$
was used.

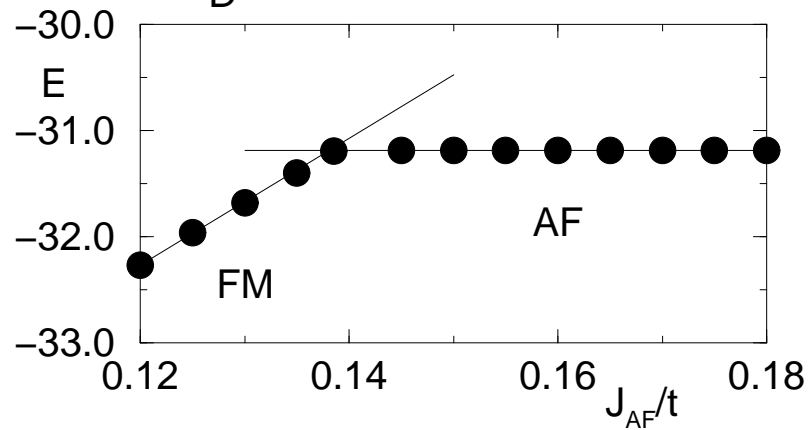
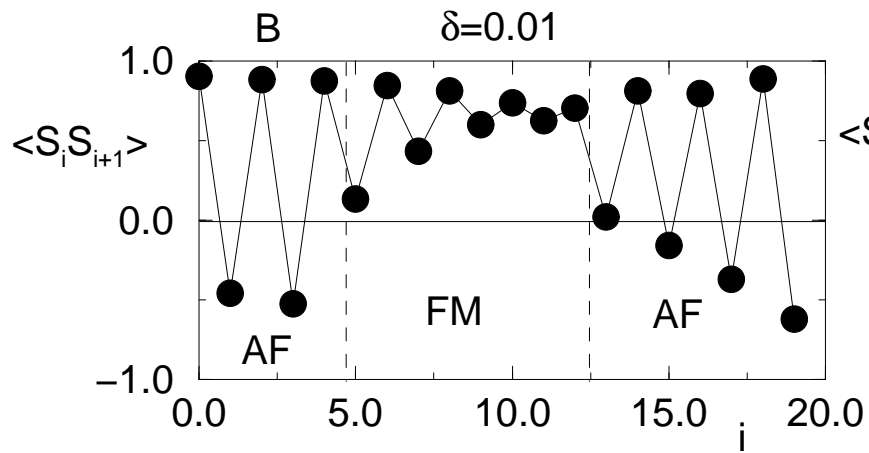
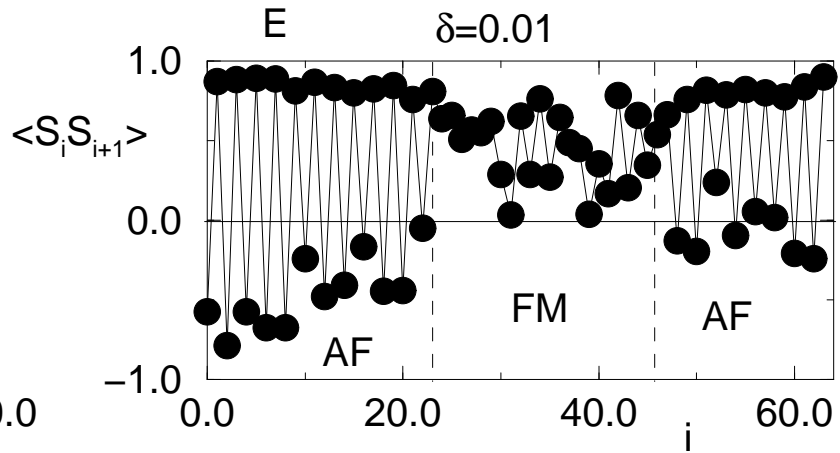
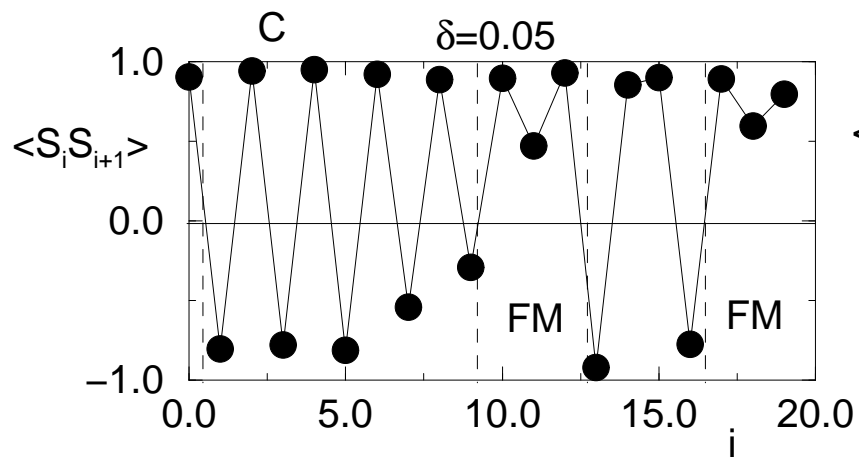
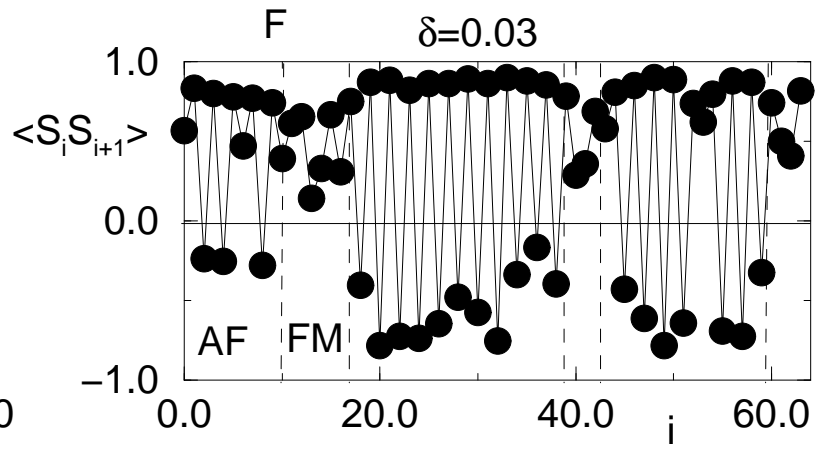
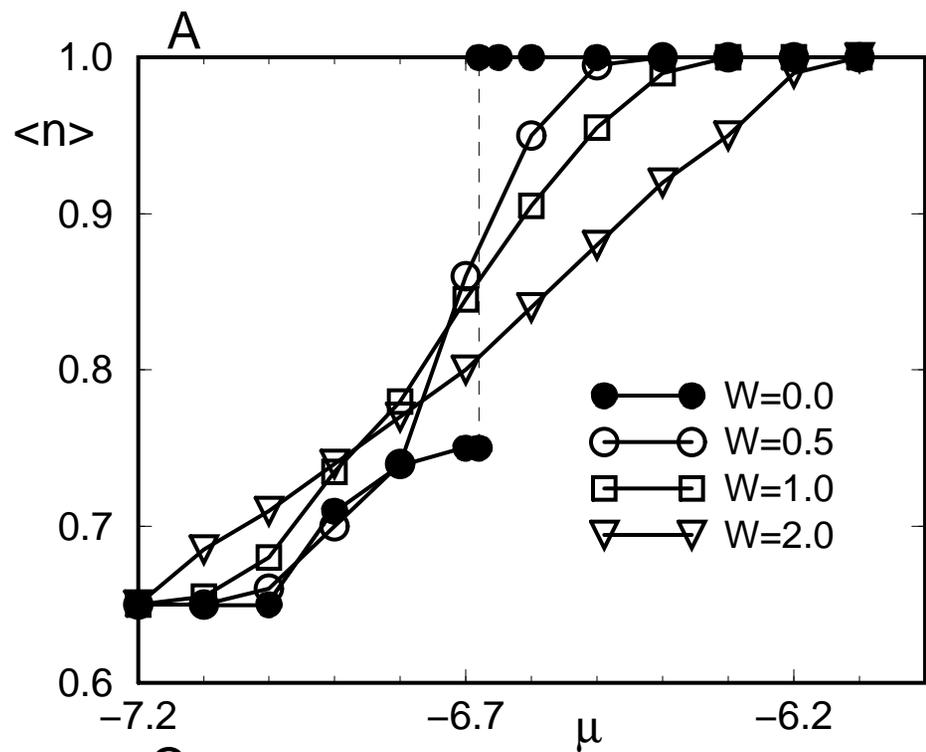
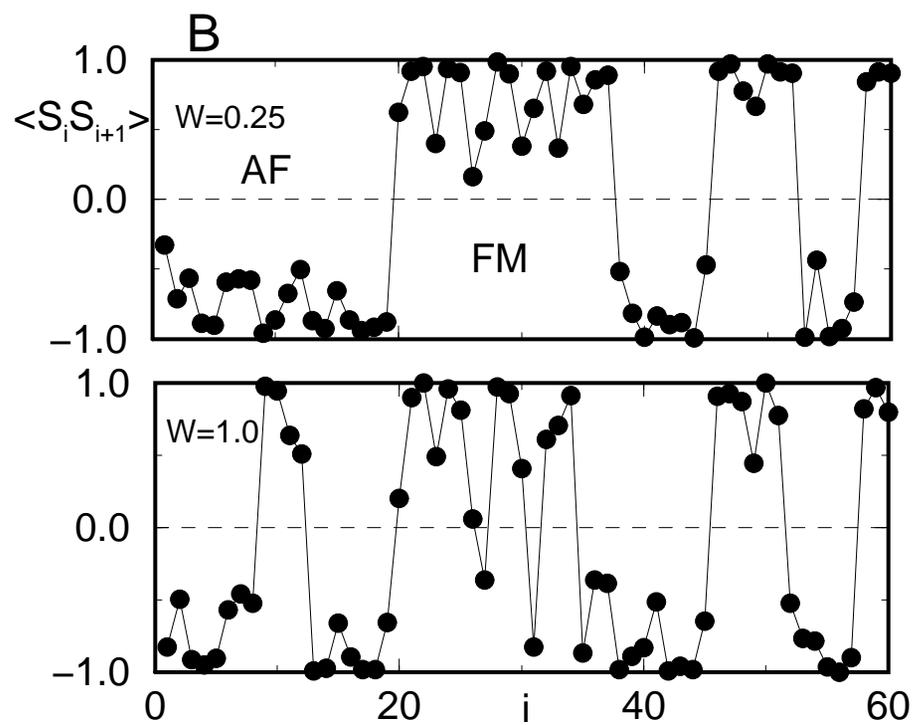
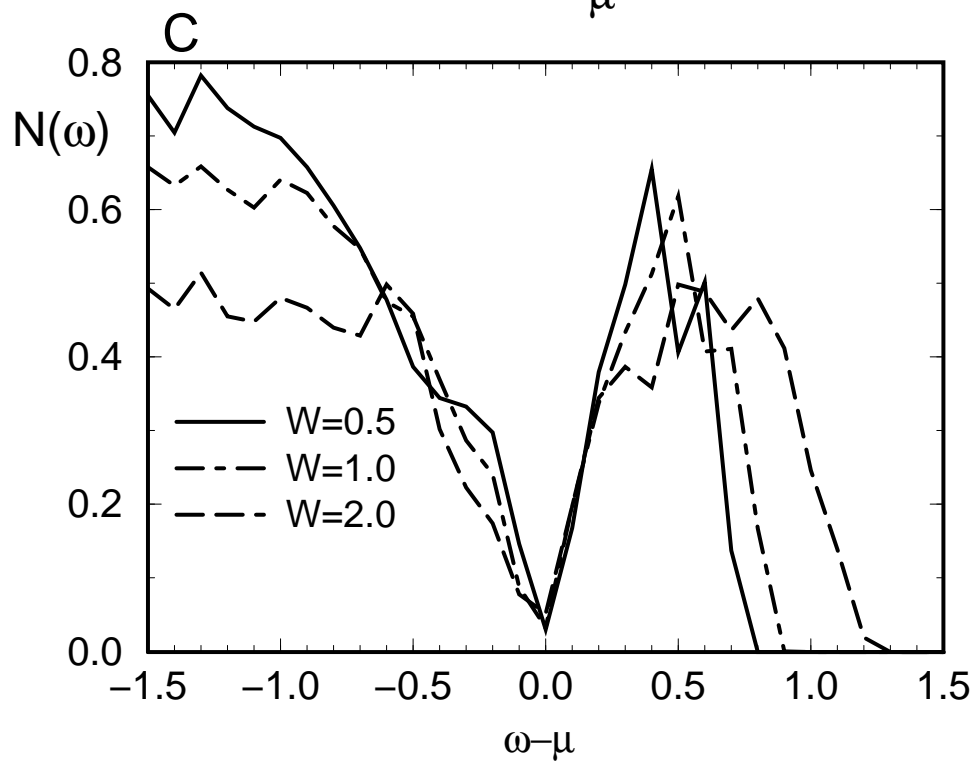
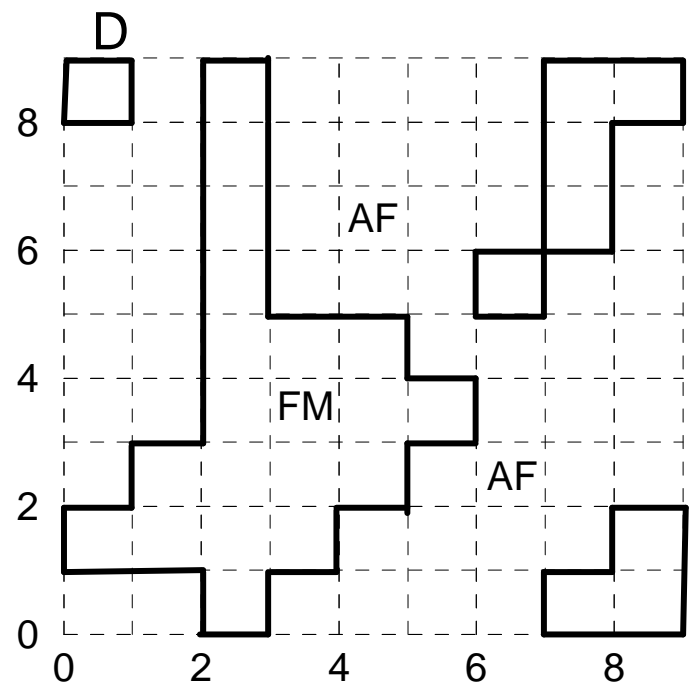
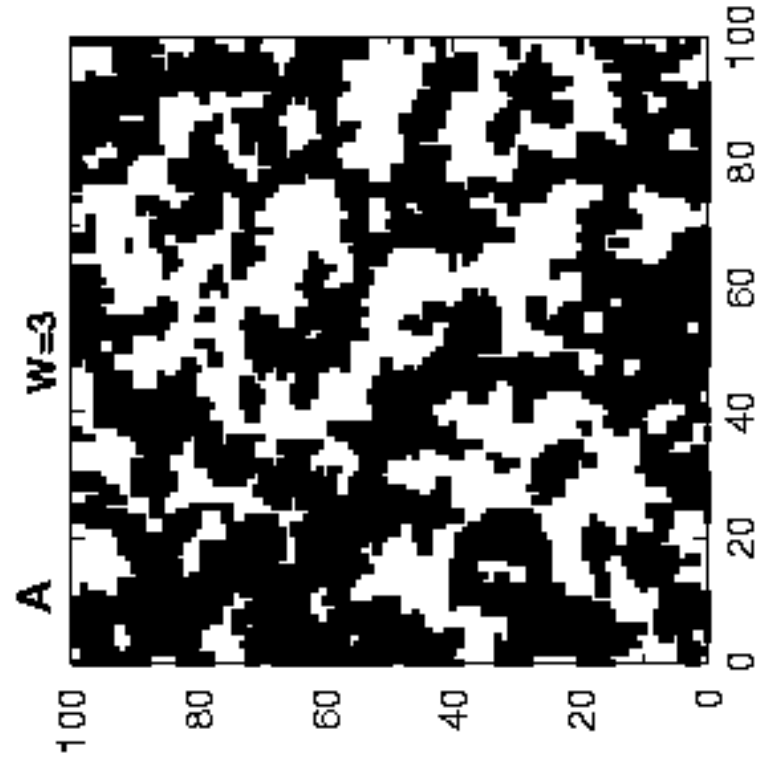
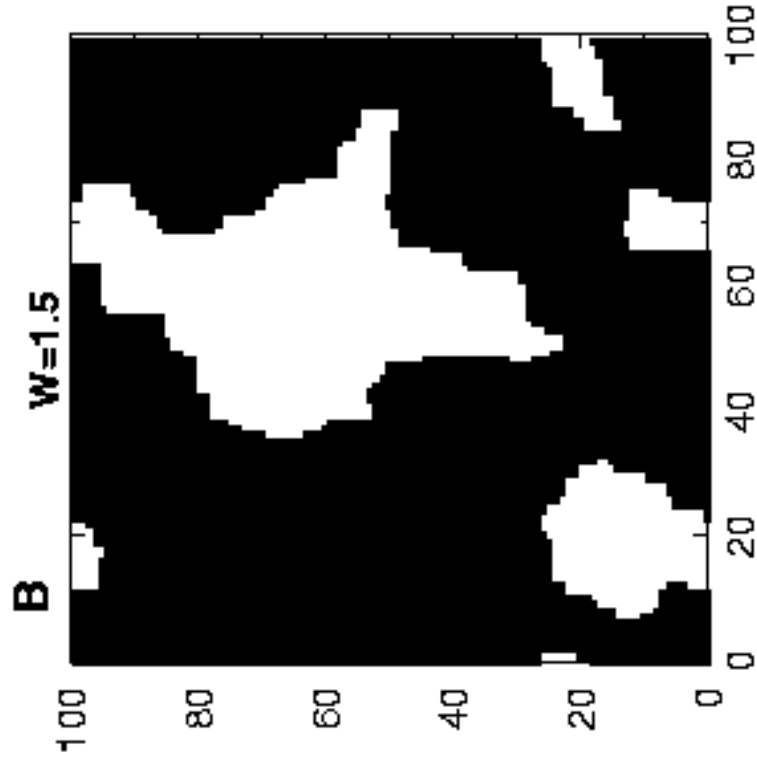
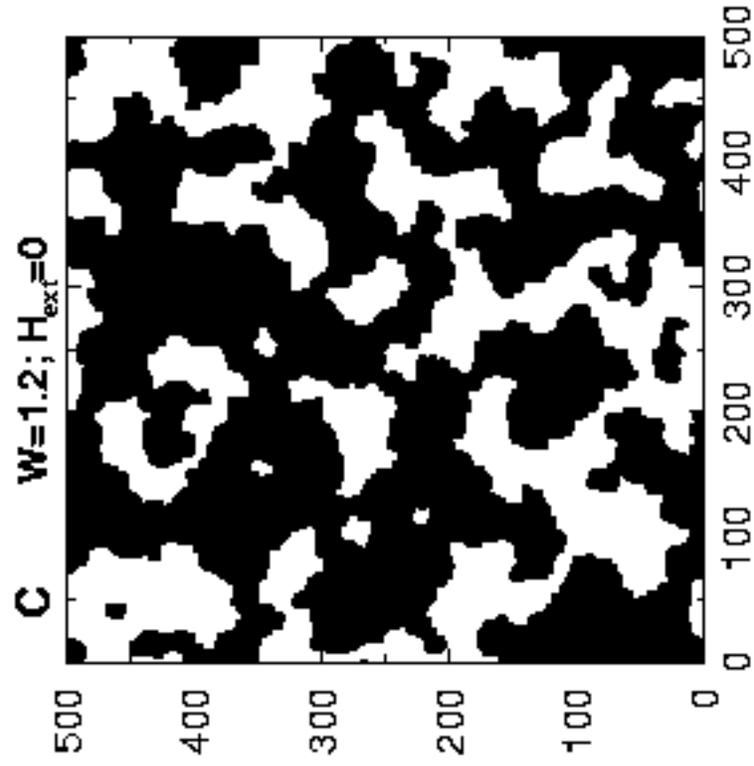
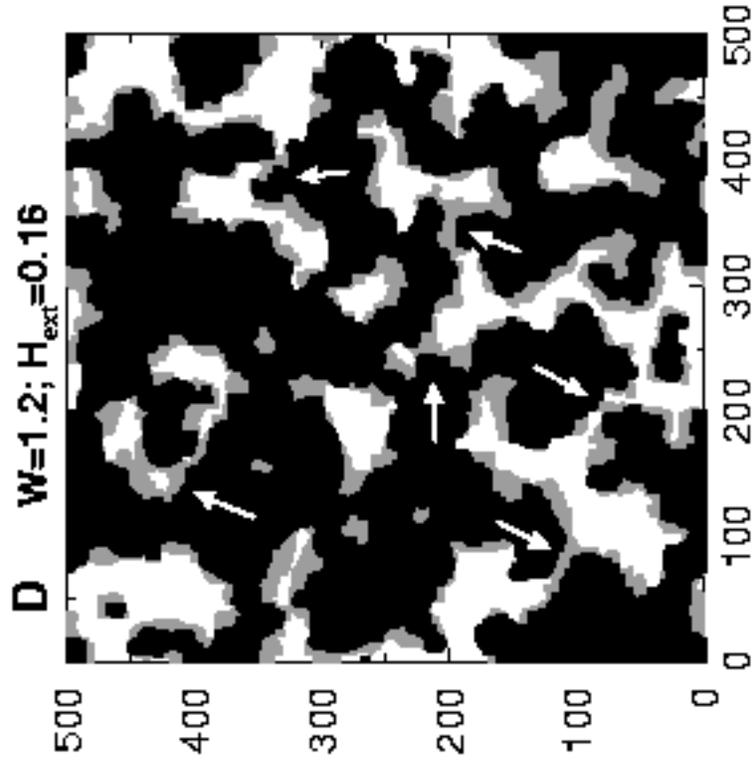
\end